\documentclass[a4paper]{aa}
\usepackage{txfonts}
\usepackage{natbib}
\usepackage{graphics, color}
\newcommand{\change}{}

\begin{document}

\def\afz{Astrofizika}%
          % Astrofizika

%\thesaurus{
%           02.16.2;
%           08.03.4
%           08.13.1;
%           08.16.5 }

\title{
%A study of photospheric and circumstellar magnetic field components in Herbig Ae stars
{\change A study of the magnetic field in the photospheric and circumstellar components
of Herbig Ae stars}\thanks{Based on observations collected at
the European Southern Observatory, Paranal, Chile (ESO programme Nos.\
072.D-0377, 073.D-0464, 074.C-0463, and 075.D-0507).}}
%Table\,\ref{t1} is only available in
%electronic form via http://www.edpsciences.org}}

\author{S. Hubrig\inst{1}
\and M.~A. Pogodin\inst{2,3}
\and R.~V. Yudin\inst{2,3}
\and M. Sch\"oller\inst{1}
\and R.~S. Schnerr\inst{4}
}

%\offprints{S. Hubrig \\
%\email{shubrig@eso.org}}

\institute{
European Southern Observatory, Casilla 19001, Santiago 19, Chile
%\email{shubrig@eso.org, mschoell@eso.org}
%\email{shubrig@eso.org}
\and
Pulkovo Observatory, Saint-Petersburg, 196140, Russia
%\email{pogodin@gao.spb.ru, ruslan61@hotmail.com}
\and
Isaac Newton Institute of Chile, Saint-Petersburg Branch, Russia
\and
Astronomical Institute ``Anton Pannekoek'', University of Amsterdam, Kruislaan 403, 1098 SJ Amsterdam, The Netherlands
}

%\date{Received 8 September 2003 / Accepted 31 March 2005}

\titlerunning{Magnetic fields of Herbig Ae stars}
\authorrunning{S.~Hubrig et~al.}

\abstract
{}
{We intend to investigate separately the photospheric and circumstellar (CS) magnetic field 
components in seven Herbig Ae stars.}
{The study is based on low-resolution ($R\sim$ 2000 and 4000)
spectropolarimetric data collected from 2003 to 2005 at the Very Large Telescope
(ESO, Chile) with the multi-mode instrument FORS1.}
{We show that the spectropolarimetric results strongly depend on the level
of CS contribution to the stellar spectra. We have improved
the determination accuracy of magnetic fields  up to the 7$\sigma$ level
in the two Herbig Ae stars HD\,139614 and HD\,144432, observed in 2005
when these objects were at a low level state of their CS activity. We have
established that at a higher level state of CS activity the polarisation
signatures are related mainly to the CS matter. The presence of CS polarisation 
signatures formed in the stellar wind supports the assumption 
that the magnetic centrifuge is a principal mechanism
of wind acceleration. 
%Further we show that the puzzling complex structure of profiles of the resonance 
%Ca\,{\sc ii} doublet in the spectrum of the Herbig Ae star HD\,190073 is related to a latitudinal
%stratification of its CS magnetic field governing the gaseous outflow.
}
{We conclude that the most effective way to investigate the magnetism of
Herbig Ae stars is to monitor their spectropolarimetric behaviour at different
states of CS activity. Obviously, higher resolution
spectropolarimetric observations would extend the sample of spectral lines to be
used for the measurements of magnetic fields at different levels in the stellar
atmosphere and CS envelope. Such observations will give a more complete insight into 
the magnetic topology in Herbig Ae stars.}

\keywords{stars: pre-main sequence --- polarisation ---
stars: magnetic fields --- stars: circumstellar matter }

\maketitle

\section{Introduction}\label{int}

The Herbig Ae/Be stars (HAEBEs) are 
pre-main sequence (PMS) objects of masses of 1.5 to 10\,M$_{\sun}$
\citep{herb,fm,the}.
In their spectra, they frequently reveal evidence of a complex
structure of the circumstellar (CS) material surrounding the star.
According to common views, the envelopes of HAEBEs contain an equatorial
gaseous-dusty disk and a stellar wind at higher latitudes. The matter
outflow from HAEBEs is time-variable and spatially inhomogeneous, and
signatures of a non-stable matter infall onto the star are frequently observed.
Presently, there is a clear lack in our knowledge of the
relation between these different processes.

Numerous theoretical works predict the existence of a global magnetic field
of a complex configuration formed as a result of the interaction between the
star and the accretion disk. Very likely, such a field is  responsible
for all observational peculiarities of HAEBEs. For this reason, an
investigation of the strength and the topology of the magnetic field can
provide important information on the origin of activity in HAEBEs.

Over an extended period of years all attempts to obtain reliable direct
measurements of magnetic fields of HAEBEs have been rather unsuccessful. On the
one hand, the lack of field detections could be plausibly explained by the
weakness of the magnetic fields in HAEBEs with strengths considerably
lower than those in magnetic A-type stars or low-mass PMS T\,Tauri objects.
In addition, the accuracy of all previous measurements was not good enough
to detect such weak fields \citep{cat93,gg}. On the other hand, the 
{\change observational strategy only allowed} to determine the average value of the magnetic
field using a large number of different spectral lines
\citep{gg}. Due to the {\change expected} complex spatial structure of magnetic fields of
HAEBEs, the measured magnetic field strength is anticipated to strongly
depend on individual spectral lines used for the field determination. The measured
magnetic field is expected to {\change vary significantly} from the stellar atmosphere to
different CS regions, even with a change of polarity.
Undoubtedly, any attempt to detect and to study
fields in HAEBEs must aim at achieving a considerably better accuracy of
measurements and at applying a specific {\change data analysis,
which we describe in the following sections}.

The first detection of a magnetic field in a Herbig Ae star
was obtained by \citet{don}. A marginal circular polarisation
signature indicating the existence of a weak magnetic field (about 50\,G)
was observed in metal lines of HD\,104237. However, numerous observations 
carried out in the last years in various wavelength regions revealed that 
HD\,104237 is a multiple system containing in addition to a Herbig A4-A7e
component four additional T\,Tauri-type components. The most massive companion 
has a bolometric luminosity that is only a factor of 10 less than the Herbig Ae primary
\citep{f,boem}.
Therefore, the possibility that the
detected magnetic field of HD\,104237 is actually related to the T\,Tauri-type
components \citep{tau} can not be ruled out.

Quite recently, definite evidence for the presence of magnetic fields
of the order of about 100\,G has been presented for several Herbig Ae stars
by \citet{h4,h6}. For the measurements of the magnetic fields we used
spectropolarimetric observations obtained with the 8\,m VLT\,+\,FORS1 in 
the years 2003 -- 2005.
The accuracy of the field determinations using 
exclusively Balmer lines in the spectropolarimetric
data obtained at a spectral resolving power $R\sim4000$ was about 30\,G. 
In these studies we reported 3$\sigma$ detections for three out of seven
Herbig Ae stars studied. 
\citet{wade} carried out a
re-analysis of the observational data for HD\,139614 obtained by
Hubrig et~al.\ in 2003.
%\citet{h4}. 
They confirmed the detection of the mean longitudinal field
in Balmer lines by measuring
a field $\langle$$B_z$$\rangle$\,=\,$-320\pm$75\,G, which is similar to our former
result $\langle$$B_z$$\rangle$\,=\,$-450\pm$93\,G.
Curiously, the magnetic field was not detected when they
also included numerous metal lines in their analysis.
%tried to measure the whole spectrum including numerous metal lines.
%($\langle$$B_z$$\rangle$\,=\,$-73\pm$40\,G). 
The presence of a weak magnetic field in HD\,139614 has been confirmed by 
our more recent spectropolarimetric observations in 2005 
\citep[$\langle$$B_z$$\rangle$\,= $-116\pm$34\,G;][]{h6}.
%($\langle$$B_z$$\rangle$\,= $116\pm$34\,G; \citet{h6}).
Three weeks after our observations \citet{wade} used ESPaDOnS at the CFHT
to obtain two additional spectropolarimetric observations of HD\,139614.
For the measurements of the magnetic field the authors utilized metal lines using 
the least-squares deconvolution method \citep[LSD;][]{don}. 
%least-squares deconvolution (LSD; \citep{don}). 
No Zeeman signature has been detected in their mean circular 
polarisation profiles (LSD Stokes~V profiles).
%We suggest that the non-detection of the magnetic field in HD\,139614 
%by \citet{wade} is related to
%the employment in the measurement procedure an inappropriate line list which included
%not only photospheric absorption lines but also the spectral lines showing emission and 
%P-Cygni profiles from the wind and/or circumstellar environment.
The non-detection  of the magnetic field in HD\,139614 
%by \citet{wade} 
is probably
related to an inappropriate line list used by these authors which included not only 
photospheric absorption lines but also spectral lines showing emission and
P\,Cygni profiles from the wind and/or circumstellar environment. 
{\change Longitudinal magnetic field measurements are only sensitive to
the line-of-sight component of the magnetic field, $\langle$$B_z$$\rangle$. This method is
most effective when lines form over a volume where the field varies slowly
with radius, such as an atmosphere where scale heights set the scale for
line formation. The resulting $\langle$$B_z$$\rangle$ is the average of the line of sight
component of the field over the stellar hemisphere visible at the time of
observations, weighted by the local emergent spectral line intensity.
%the longitudinal Zeeman effect
%measures the net magnetic flux of the line-of-sight, yielding $\langle$$B_z$$\rangle$, the
%method is best when lines form over a volume where the field is slowly
%varying with radius, such as an atmosphere where scale heights set the
%scale for line formation. $\langle$$B_z$$\rangle$ is the average of the component 
%of the field parallel to the line of sight
%over the stellar hemisphere visible at the time of observations,
%weighted by the local emergent spectral line intensity. 
In a circumstellar environment, such as a wind or an accretion disk,
lines may form over a relatively large volume, and the field topology may
be complex not only in latitude and azimuth, but in radius as well.
Thus, for complex magnetohydrodynamical flows in the presence of photospheric and 
circumstellar fields of comparable strength it is quite easy to drive the net 
line-of-sight magnetic flux to near zero values. 
Other possibilities to explain the discrepancy between the different authors could 
be the expected variability of the field strength resulting from the 
variable CS contribution or the rotational modulation of the magnetic field
provided this star represents an inclined magnetic rotator.
Even in the case where the photospheric fields are stronger than circumstellar fields,
the presence of wind line emission would produce diminishing of the circular polarization signal,
i.e.\ polarimetric dilution caused by contributing unpolarized emission. 
If both photospheric and
circumstellar contributions are present and an unparsed ensemble of lines is
used, null detections may result because of the issues 
described here.}
%the presence of the magnetic field in HD\,139614 was confirmed by
%our new VLT observations: MJD\,53405, $\langle$$B_z$$\rangle$\,=
%116$\pm$34\,G \citep{h6} but it was not confirmed by high-resolution
%spectropolarimetric measurements with the CFHT\,+\,ESPaDOnS obtained
%three weeks later \citep{wade}: MJD\,53423,
%$\langle$$B_z$$\rangle$\,=\,-20$\pm$25\,G, and MJD\,53424,
%$\langle$$B_z$$\rangle$\,=\,-35$\pm$25\,G.  We propose that these
%discordances can be connected with: ~ a) different field parameters
%diagnosed in lines originating in different regions, and ~ b)
%time-variability of field parameters which can be a result of a real
%change of the CS influence as well as of the rotational modulation
%provided the object represents an inclined magnetic rotator. In
%one's turn, Further, using the CFHT\,+\,ESPaDOnS \citet{wade} have detected 
%magnetic fields in two more
%Herbig Ae stars: \object{HD\,101412},
%$\langle$$B_z$$\rangle$\,=\,+430$\pm$75\,G ~ and \object{V380\,Ori},
%$\langle$$B_z$$\rangle$\,=\,-460$\pm$70\,G.

Our previous magnetic field measurements of seven Herbig 
Ae stars have been carried  out
exclusively in the broad hydrogen lines \citep{h4,h6}.
In the new analysis we have used the full spectrum, excluding only the regions
which are contaminated by the circumstellar environment.
Below, we report our results of analysing the magnetic fields in both atmosphere
and circumstellar environment of this Herbig star sample. We also included 
in this analysis two new spectropolarimetric observations of the Herbig 
Ae star HD\,190073 
and one observation of the Herbig Ae star HD\,144668, all obtained in 2005. 
Considerable attention has been paid to 
the presentation of the circular polarisation spectra to demonstrate
the Zeeman features indicating the presence of magnetic fields.
 
%With the above reasoning we have re-analysed the observational data for 7
%objects of our programme reported in \citet{h4,h6} separately for spectral
%lines forming in the stellar atmosphere and in the CS material. The results of
%the analysis are presented in this paper. Understanding that an estimation
%obtained with the accuracy at the 3$\sigma$ level is yet not be a convincing
%evidence for the magnetic field existence, we have received sufficient
%attention to graphical illustrations of the circular polarisation spectra to
%demonstrate more clearly the observed polarisation features indicating the
%presence of magnetic fields.

\section {Program targets and strategy of analysis}\label{progr}

To analyse photospheric and circumstellar magnetic field components in Herbig Ae
stars we used spectropolarimetric observations of seven Herbig Ae stars 
obtained in 2003 -- 2005 with the multi-mode instrument
FORS\,1 installed at the 8\,m-KUEYEN telescope at the VLT (ESO, Chile).
The circular polarisation measurements
carried out in the wings of Balmer lines have already been reported by  \citet{h4,h6}.
Two additional observations of the Herbig Ae star HD\,190073 have been obtained at the end of May 
and in August 2005. One additional {\change observation} of the Herbig Ae star HD\,144668 {\change was}
obtained in April 2005. These have also been included in the {\change present} analysis. 

FORS\,1 is equipped with polarisation analyzing optics comprising super-achromatic half-wave and 
quarter-wave phase retarder plates and a Wollaston prism with a beam divergence 
of 22$^{\prime\prime}$ in standard resolution mode. We used the GRISM 600B to cover all H 
Balmer lines 
from H$\beta$ to the Balmer jump and GRISM 1200g to cover the H Balmer lines from H$\beta$ to H8. 
The spectral resolution of the FORS\,1 spectra obtained with GRISM 600B is about 2000 whereas 
GRISM 1200g allows us to achieve a spectral resolving power of $R\sim4000$ with the narrowest 
available slit width of 0\farcs4.
The spectral regions covered by GRISM 600B and GRISM 1200g are respectively 
$\lambda\lambda$\,3700--5900\,\AA{} and $\lambda\lambda$\,3900--5000\,\AA{}.
A detailed description of the assessment of the longitudinal magnetic field measurements 
using the FORS\,1 spectra is presented in our previous papers \citep{h4,h6} and in
references therein.

The technique applied for measuring magnetic fields in stellar atmospheres was 
developed by \citet*{al}, allowing the determination of the mean longitudinal magnetic 
field $\langle$$B_z$$\rangle$ of the star.
%$\langle$$B_z$$\rangle$ is the average of the component of the field parallel to the line of sight
%over the stellar hemisphere visible at the time of observations,
%weighted by the local emergent spectral line intensity. 
It works well if the line profiles are 
axially symmetric and if the Zeeman splitting is small compared to the line intrinsic
broadening (so-called weak-field regime).
However, their technique can not be applied to spectral lines originating in the CS material:
Profiles of these spectral lines are mainly determined by the velocity
field in the CS gas and are seldom axially symmetric. The motion of
CS gas is expected to be strongly governed by a magnetic field, especially in
the regions of the stellar wind, which is generally assumed to be accelerated by
a magnetic centrifuge. Therefore, different parts of individual line profiles can
diagnose CS regions with very different strengths of the magnetic field.
Nevertheless, circular polarisation signatures observed at
the wavelengths corresponding to clearly identified CS lines allow to
study a magnetic field in the region of the line
formation.
%If the observed circular polarisation signature is
%rather conspicuous, one can assume the presence of a magnetic field.
If such circular polarisation features are indeed observed, one can infer
the presence of a magnetic field.
However, the numerical determination of
$\langle$$B_z$$\rangle$ in the CS region can be carried out
only by using a special method.

The goals of our study are as follows: ~ a) to analyse the 
spectropolarimetric data of seven program stars and to separate  photospheric
and CS components of all spectral lines; ~ b) to re-measure the photospheric magnetic
fields of Herbig Ae stars without inclusion of
spectral regions affected by the CS matter; ~ c) to derive information on
the presence of magnetic fields in the CS environment.
{\change The idea to detect the Zeeman effect in the CS environment
is rather new.
The first theoretical assessment of wind signatures in Stokes V profiles has been 
presented by \citet{ign}.}

To determine the spectral region affected by the CS matter, we have compared
for each star the observed 
spectra with synthetic photospheric spectra computed for models of stellar atmospheres with
corresponding stellar parameters $T_{\rm eff}$ and log\,$g$. The synthetic photospheric
spectra were calculated with the computer code SYNTH\,+\,ROTATE developed
by \citet{p}. Because of the rather low resolution of our spectra 
the fitting has been applied only to slopes of wide
Stark broadened wings of the Balmer H$\beta$, H$\gamma$, H$\delta$, and  H$\epsilon$ lines. 
The results of our assessment of atmospheric parameters are presented in
Table\,\ref{t1} along with the values of $v\,\sin\,i$,
%taken in part from \citet{h6} 
which were determined from high-resolution non-polarized spectra collected in recent years
with EMMI at the NTT  on La Silla and at the Crimean Observatory.

\begin{table}
\begin{center}
\caption[]{Atmospheric parameters used for the calculation of
synthetic photospheric spectra of the program targets.}
\label{t1}
\begin{tabular}{cccc}
\hline\hline\noalign{\smallskip}
 HD  &  $T_\mathrm{eff}$ (K) & log\,$g$ & $v\,\sin\,i$ \\
\noalign{\smallskip}\hline\noalign{\smallskip}
  31648    &  9250 & 3.5 & 90 \\
  38238    &  7750 & 3.5 & 100 \\
  139614   &  8250 & 4.0 & 15 \\
  144432   &  7250 & 4.0 & 70 \\
  144668   &  7500 & 4.0 & 100 \\
  163296   &  9250 & 3.5 & 130 \\
  190073   &  9250 & 4.0 & 12 \\
\noalign{\smallskip}\hline
\end{tabular}
\end{center}
\end{table}

\begin{table*}
\begin{center}
\caption[]{Results of the magnetic field measurements for each group.
We give the HD numbers and modified Julian dates of the observations followed by the resolution of the spectropolarimetric 
data.
$\langle$$B_z$$\rangle$ ``previous'' refers {\change to the analysis} in \citet{h6},
$\langle$$B_z$$\rangle$ ``new'' is described in Sect.~\ref{res}.
$(+)$ means that the stellar magnetic field is detected with a high level
of confidence, and $(?)$ indicates that although the field is detected, it
is possibly distorted by the CS material.
The last column contains lines for which
signatures of CS magnetic field are clearly seen.}
\label{t2}
\begin{tabular}{ccccrrcc}
\hline\hline\noalign{\smallskip}
Group  &
HD &
MJD &
$R$ &
\multicolumn{1}{c}{$\langle$$B_z$$\rangle$} &
\multicolumn{1}{c}{$\langle$$B_z$$\rangle$} &
Stellar &
CS \\
   &   & & &
\multicolumn{1}{c}{previous} &
\multicolumn{1}{c}{new} &
magnetic field & magnetic field \\
\noalign{\smallskip}\hline\noalign{\smallskip}
 I   & 139614 & 53405.37 & 4000 & $-$116$\pm$34\,G & $-$93$\pm$~~14\,G & + &   \\
 & 144432 & 53447.35 & 4000 & $-$119$\pm$38\,G & $-$111$\pm$~~16\,G & + &
 H$\beta$-H$\epsilon$, Ca\,{\sc ii} H+K \\
\noalign{\smallskip}\hline
 II    & 38238 & 53249.37 & 2000 & $-$115$\pm$67\,G & +13$\pm$~~36\,G &   &
 H$\beta$-H$\epsilon$, Ca\,{\sc ii} H+K \\
   & 139614 & 52904.04 & 2000 & $-$450$\pm$93\,G & $-$112$\pm$~~36\,G & ? &
  H$\beta$-H$\epsilon$, Ca\,{\sc ii} H+K  \\
 & 144432 & 52900.99 & 2000 & $-$94$\pm$60\,G & +32$\pm$~~37\,G &   &
 H$\beta$-H$\epsilon$, Ca\,{\sc ii} H+K \\
  & 144668 & 52901.01 & 2000 & $-$118$\pm$48\,G & +166$\pm$~~40\,G & ? &
 H$\beta$-H${10}$, Ca\,{\sc ii} H+K \\
  & 144668 & 53120.25 & 2000 & $-$107$\pm$40\,G & $-$75$\pm$~~29\,G &  &
 H$\beta$-H${10}$, Ca\,{\sc ii} H+K \\
  & 144668 &  53461.30    & 4000 &  \multicolumn{1}{c}{---}   & +195$\pm$121\,G &  &
 H$\beta$-H${10}$, Ca\,{\sc ii} H+K \\
\noalign{\smallskip}\hline
 III    & 31648 & 53296.35 & 4000 & +87$\pm$22\,G & +73$\pm$~~32\,G &   &
 H$\beta$-H$\epsilon$, Ca\,{\sc ii} H+K \\
  & 163296 & 53279.00 & 4000 & $-$57$\pm$33\,G & $-$25$\pm$~~27\,G &  &
 Ca\,{\sc ii} H+K \\
 & 190073 & 53514.36 & 4000 & \multicolumn{1}{c}{---} & +21$\pm$~~12\,G &  &
 Ca\,{\sc ii} H+K \\
  & 190073 & 53519.38 & 4000 & +84$\pm$30\,G & +14$\pm$~~22\,G &  &
 Ca\,{\sc ii} H+K \\
  & 190073 & 53596.15 & 4000 & \multicolumn{1}{c}{---} & $-$51$\pm$~~14\,G & ?  &
 Ca\,{\sc ii} H+K \\
\noalign{\smallskip}\hline
\end{tabular}
%\parbox{\hsize}{
%{\footnotesize Notes: $\langle$$B_z$$\rangle$ (previous) refers to
%\citet{h6}; $\langle$$B_z$$\rangle$ (new) is described in this
%paper; $(+)$ means that the stellar magnetic field is present at high level
%of significance, and $(?)$ -- indicates that although the field is detected, it
%is possibly distorted by the CS material ; the last column contains lines, for which
%signatures of CS magnetic field are clearly seen.}}
\end{center}
\end{table*}

\section{Results}\label{res}

We suggest
that the observations of the studied Herbig Ae stars can be divided into three groups. The first
group contains objects with photospheric magnetic fields which have been
detected at a high confidence level of
about 7$\sigma$. The objects with only weak circular polarisation signatures
which however can not be determined with sufficient precision
constitute the second group. Objects with clear circular polarisation signatures
mainly of CS origin have been distinguished as a third group.
The magnetic field has been remeasured exclusively in spectral regions not
contaminated by the CS material.

The results of our analysis are summarised in Table\,\ref{t2}.
{\change Each row corresponds to one measurement.
In the first column we identify the group a measurement falls in.
The second column gives the name of the target, the third column the modified Julian date
of the observation, and the fourth column the spectral resolution of the measurement.
In column five we list the value for $\langle$$B_z$$\rangle$ from the analysis in \citet{h6},
where we used all Balmer lines, except for HD\,190073, where we used only \ion{Ca}{ii} H+K.
In column six we list the value for $\langle$$B_z$$\rangle$ from the analysis reported in this
paper, using the same data.
Here, we have taken the full spectrum, except for the Balmer lines and the \ion{Ca}{ii} lines, where 
we detect contamination of the lines by the CS environment.
In case we detect the stellar magnetic field with a high confidence level,
we indicate this in column seven by a (+), in case we find a field but believe it is distorted
by the CS material, we indicate this in column seven by a (?).
In the last column we list the lines for which signatures of a CS magnetic field are clearly seen.}
In the following we discuss the results obtained for each group,

\subsection{Group I}\label{i}

\subsubsection{HD\,139614 (observed in 2005)}\label{ia}

\begin{figure}
\centering
%\sidecaption
\resizebox{.45\textwidth}{!}{\includegraphics{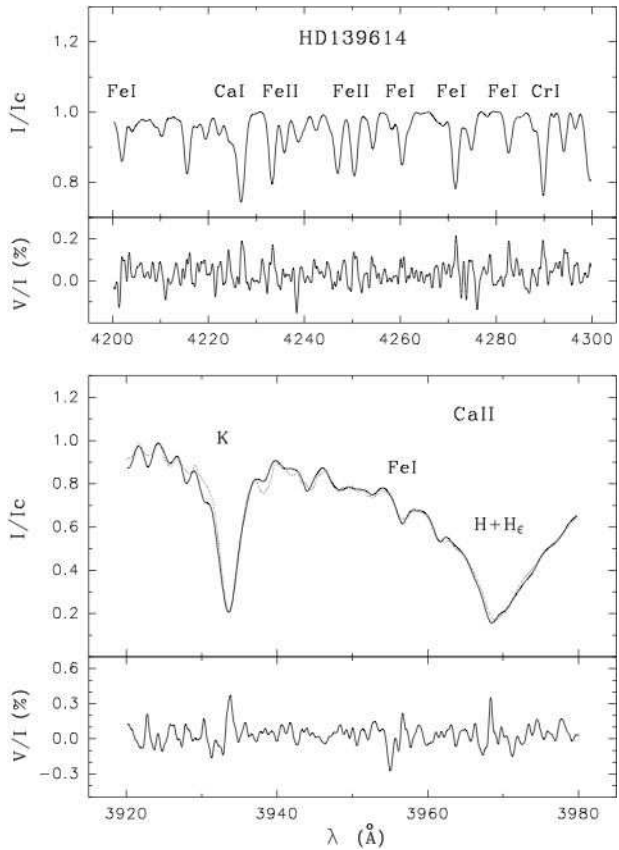}}
\caption[]{Stokes~I and V spectra of HD\,139614 in two different spectral regions,
observed in 2005.
The dotted vertical lines in the upper panel indicate polarisation signatures corresponding to
positions of stellar atmospheric lines.
The synthetic photospheric spectrum in the lower panel is shown by the dotted line.}
\label{f1}
\end{figure}

The comparison of the spectrum of HD\,139614 observed in 2005 with the synthetic photospheric
spectrum reveals that the spectrum is mainly
of photospheric origin. The only exceptions are the first members of the
Balmer series. Broad profiles of these lines are overlapped by
low-intensity central emissions whose strength drops gradually from
H$\beta$ to H$\delta$. The spectrum of the object in the blue spectral region
$\lambda\lambda$\,4200--4300\,\AA{} contains plenty of metal lines. The
Stokes~V spectrum in this region shows a noticeable correlation between the 
wavelength of many absorption lines and circular polarisation signatures (Fig.\,\ref{f1}, top).
The most prominent polarisation signatures are observed in the H and K components
of the resonance Ca\,{\sc ii} doublet as well as in the Fe\,{\sc i} blend
at $\lambda$\,3956.8\,\AA{} (Fig.\,\ref{f1}, bottom).

No polarisation signatures are observed in the H$\beta$, H$\gamma$, and H$\delta$ lines.
We explain this fact by a significant contribution of CS material distorting 
the circular polarisation signatures of the atmospheric component. Excluding these
distorted spectral regions from the evaluation of the magnetic field, we 
determine a very accurate value for the magnetic field in this star 
$\langle$$B_z$$\rangle$\,=\,$-93\pm$14\,G.

\subsubsection{HD\,144432 (observed in 2005)}\label{ib}

\begin{figure}
\centering
%\sidecaption
\resizebox{.45\textwidth}{!}{\includegraphics{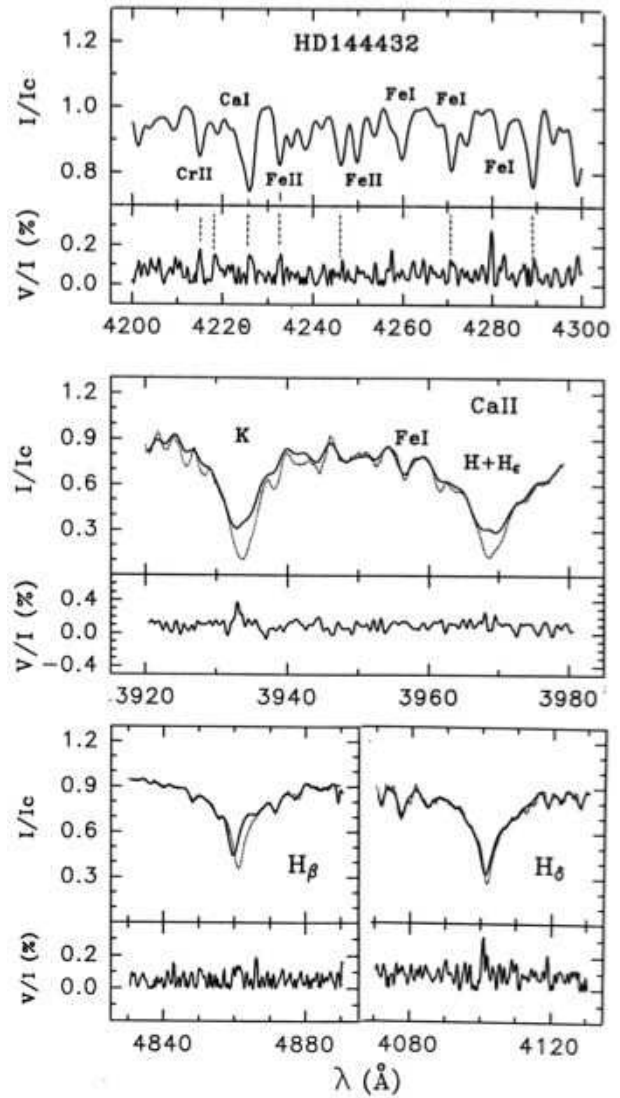}}
\caption[]{Stokes~I and V spectra of HD\,144432 in four different spectral regions,
observed in 2005.
The dotted vertical lines in the upper panel indicate polarisation signatures corresponding to
positions of stellar atmospheric lines.
The synthetic photospheric spectrum is shown in the other panels by the dotted line.}
\label{f2}
\end{figure}

The spectrum of HD\,144432 reveals signatures of a dense stellar wind,
{\change i.e.\ the wind emission is strong enough to appear in the lines.}
Blueshifted absorption components are clearly seen in H$\beta$, H$\gamma$,
H$\delta$, H$\epsilon$, and the Ca\,{\sc ii} doublet lines.
Strong circular polarisation signatures at the wavelengths of these lines
indicate that outflowing gas in the wind region is
magnetized (Fig.\,\ref{f2}). Similar to HD\,139614,
HD\,144432 shows noticeable polarisation signatures in
metal lines. Excluding the spectral regions containing the hydrogen and 
Ca\,{\sc ii} doublet lines, we determine the photospheric magnetic
field $\langle$$B_z$$\rangle$\,=\,$-111\pm$16\,G.
%No field has been detected 
%on the Stokes~V spectrum of HD\,144432 (see Table\,\ref{t2}).

\subsection{Group II}\label{ii}

\subsubsection{HD\,38238}\label{iia}

\begin{figure}
\centering
%\sidecaption
\resizebox{.45\textwidth}{!}{\includegraphics{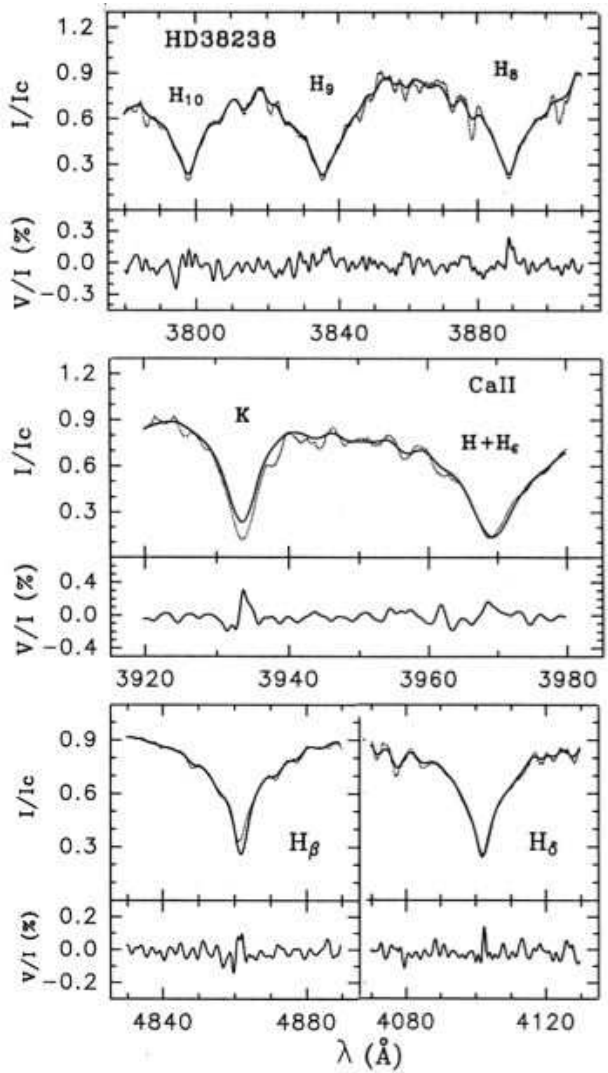}}
\caption[]{Stokes~I and V spectra of HD\,38238 in four different spectral regions.
The synthetic photospheric spectrum is shown in all panels by the dotted line.}
\label{f3}
\end{figure}

The CS magnetic field component in the  spectral lines of
HD\,38238 is only weakly present (Fig.\,\ref{f3}). Faint
circular polarisation signatures are detectable in Balmer and Ca\,{\sc ii}
lines, but they probably represent a combination of the stellar
atmospheric  and CS magnetic field signatures. No 
magnetic field has been detected in this star.

\subsubsection{HD\,139614 (observed in 2003)}\label{iib}

\begin{figure}
\centering
\begin{center}
\resizebox{.45\textwidth}{!}{\includegraphics{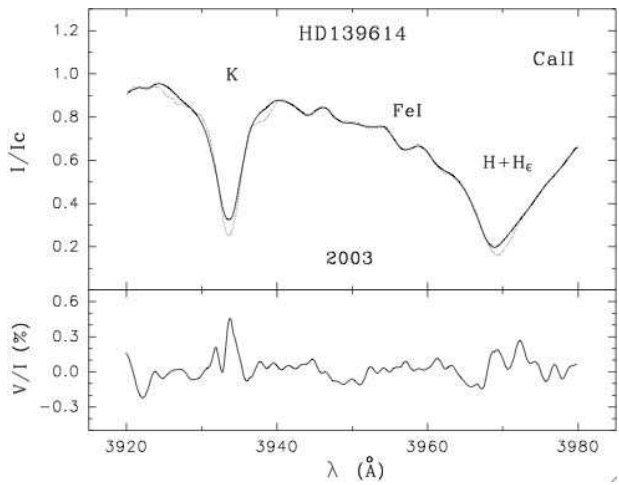}}
\end{center}
%\vskip-.4cm
%\caption[]{The H$\beta$ profiles of HD\,139614 observed in 2003 
%({\it top}) and 2005 ({\it bottom}) together with the corresponding Stokes~V spectra.
%The synthetic photospheric spectrum is shown by the dotted line.  }
\caption[]{Stokes~I and V spectra of HD\,139614 around the Ca\,{\sc ii} doublet.
The synthetic photospheric spectrum is shown by the dotted line.}
\label{f4}
\end{figure}

Previous spectropolarimetric observations of HD\,139614 carried out in
2003 \citep{h4} showed a significantly higher level of CS activity in comparison
with the 2005 data.
Fig.\,\ref{f4} presents the Stokes V spectrum of HD139614
obtained near the Ca\,{\sc ii} doublet in 2003. The lower spectral
resolution ($R\sim$ 2000) does not allow to estimate very accurately the CS
contribution to the stellar spectrum in comparison with the
synthetic spectrum of the object. Nevertheless, we believe that 
the CS contribution in 2003 was significantly higher than in 2005. The
Stokes~V spectrum reveals circular polarization signatures in the
Ca K and H components displaying double-peaked features. The feature
in the Fe\,{\sc i} blend
at $\lambda$\,3956.8\,\AA{} is clearly visible in 2005 (Fig.\,\ref{f1}),
but is absent
in 2003. Similar circular polarization signatures were also
observed in H$\beta$, H$\gamma$, and H$\delta$ lines.
%We assume that they are a result of a superposition of the stellar and
%CS components of magnetic field. Therefore,
%the significantly more intense emission
%in the centre of the H$\beta$ line in 2003. The Stokes~V spectrum reveals a circular 
%polarisation signature indicating the presence of a magnetic field of 
%negative polarity.
%The circular polarisation signatures were observed in Balmer lines only in 2003. 
Thus, we assume that the magnetic 
field measured in 2003 \citep{h4} is mainly of CS origin.
Our re-analysis of these 
observations excluding the spectral regions containing hydrogen lines leads to 
a magnetic field detection at the 3$\sigma$ level (see Table\,\ref{t2}).

\subsubsection{HD\,144432 (observed in 2003)}\label{iic}

\begin{figure}
\centering
%\sidecaption
\resizebox{.45\textwidth}{!}{\includegraphics{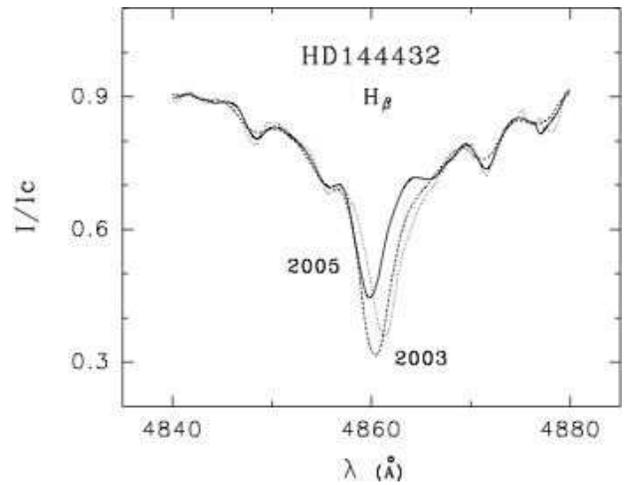}}
\caption[]{The H$\beta$ profiles of HD\,144432 observed in 2003
(dashed line) and 2005 (solid line). The synthetic photospheric spectrum is shown by
the dotted line.
}
\label{f5}
\end{figure}

Fig.\,\ref{f5} demonstrates a notable difference in the H$\beta$ profiles 
observed in 2003 and 2005. In 2003 this line
was significantly deeper and displayed no emission. In
contrast to the observations in 2005, no photospheric magnetic field has been 
detected in 2003, thus a CS contribution to the photospheric spectrum was likely 
stronger during that measurement. It is possible that a larger amount of dense outflowing
gas was accumulated around the star. As a
consequence, the circular polarisation signatures of the stellar magnetic field
which are clearly visible in 2005, were likely distorted by CS material in 
the spectrum observed in 2003.

\subsubsection{HD\,144668}\label{iid}

\begin{figure}
\centering
%\sidecaption
\resizebox{.45\textwidth}{!}{\includegraphics{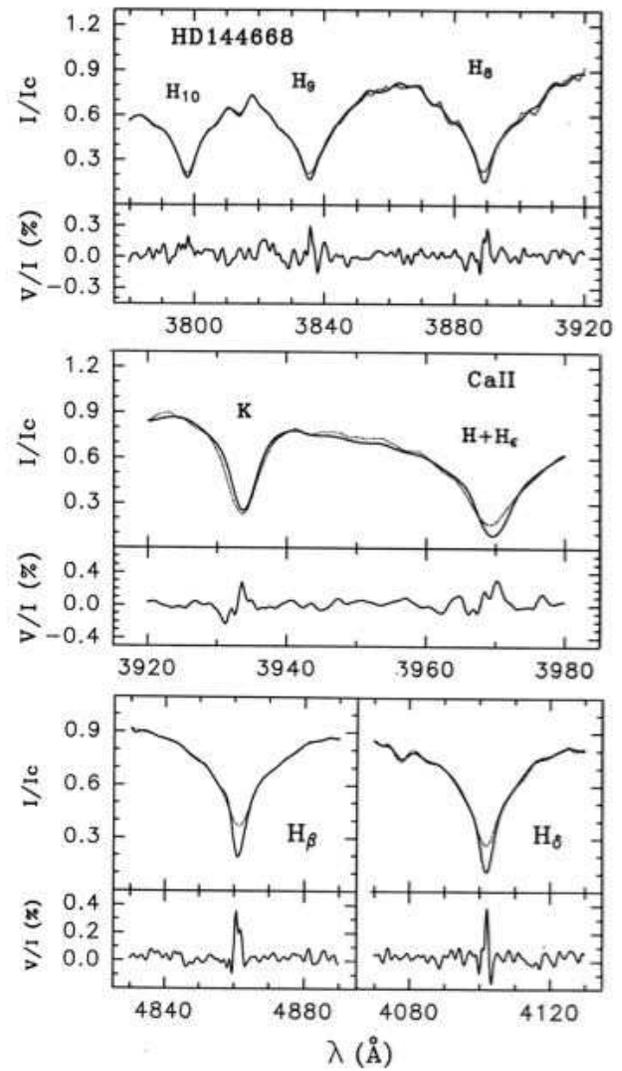}}
\caption[]{
%The same caption as in Fig.\,\ref{f1} but for
HD\,144668 on MJD\,53120.25 (see caption of Fig.\,\ref{f3}).
}
\label{f6}
\end{figure}

As in the case of HD\,38238, the first two spectropolarimetric 
observations of HD\,144668 have been carried out at lower spectral resolution
($R\sim2000$). The Stokes~V spectrum of this object as well as the
contribution of the CS matter to the  spectrum are also similar to
those of HD\,38238 (Fig.\,\ref{f6}). We were able to measure
a weak photospheric magnetic field only for 
MJD\,52901.01 at a significance level of 4$\sigma$ 
($\langle$$B_z$$\rangle$\,=\,166$\pm$40\,G).

\subsection{Group III}\label{iii}

\subsubsection{HD\,31648}\label{iiia}

\begin{figure}
\centering
%\sidecaption
\resizebox{.45\textwidth}{!}{\includegraphics{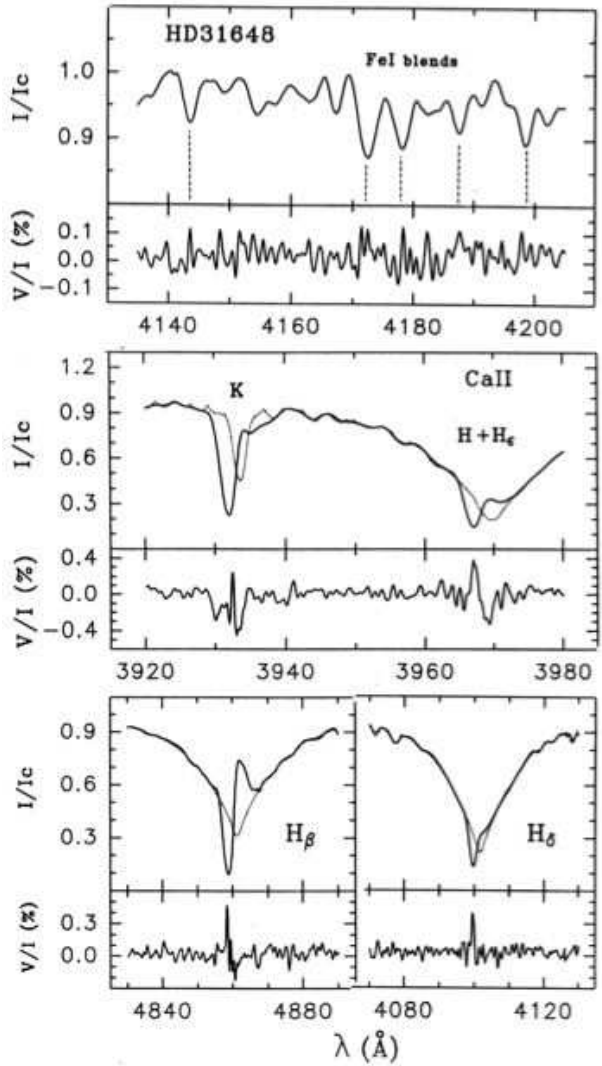}}
\caption[]{
%The same caption as in Fig.\,\ref{f1} but for HD\,31648.
HD\,31648 (see caption of Fig.\,\ref{f2}).
}
\label{f7}
\end{figure}

HD\,31648 is one of the hottest targets of the programme,
demonstrating notable emissions in H$\beta$, H$\gamma$, and H$\delta$ lines which
indicate the presence of a significant stellar wind. The profiles of
these lines are of P\,Cyg-type\,{\sc ii} \citep[see][]{beals} with very deep
blueshifted absorptions (Fig.\,\ref{f7}). 
Strong circular polarisation signatures associated
with the CS component implicate the presence of a considerable magnetic field
in the outflowing gas. Another evidence for the existence of the field is provided by
the presence of notable polarisation signatures in the
Ca\,{\sc ii} lines. The K profile of the doublet is of complex type and 
consists of two components: a blueshifted deep absorption and a red
narrow one. The Ca\,{\sc ii} lines in the spectrum of HD\,31648 are
very likely formed at the base of the stellar wind as well as in the
accretion gaseous flow. The presence of weak polarisation features is 
visible in metallic lines. We conclude that a significant magnetic field
is present in HD\,31648, but it is mostly of CS origin.

\subsubsection{HD\,163296}\label{iiib}

\begin{figure}
\centering
%\sidecaption
\resizebox{.45\textwidth}{!}{\includegraphics{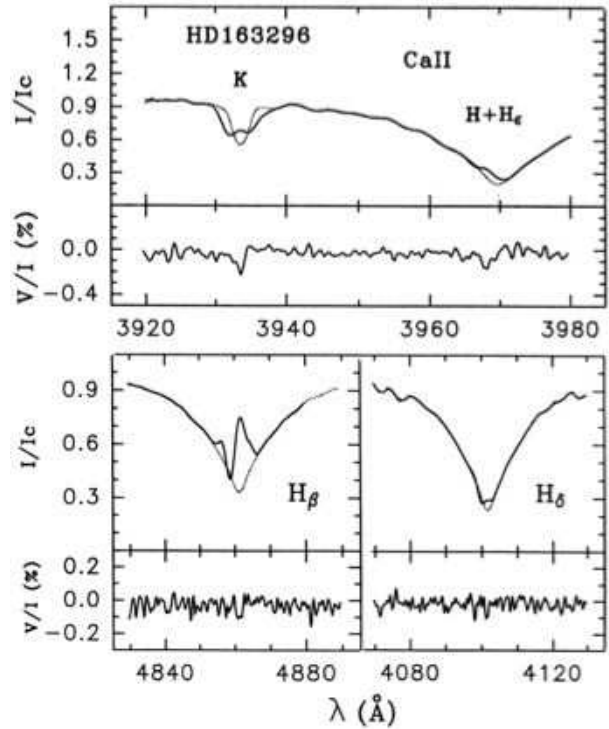}}
%\caption[]{The same caption as in Fig.\,\ref{f1} but for HD\,163296.
\caption[]{Stokes~I and V spectra of HD\,163296 in three different spectral regions.
The synthetic photospheric spectrum is shown in all panels by the dotted line.}
\label{f8}
\end{figure}

The Stokes~I spectrum of HD\,163296 exhibits emission in Balmer lines, similar to that 
of HD\,31648.
However, the Balmer line profiles are of 
P\,Cyg-type\,{\sc iii} with a secondary blueshifted emission
component and a shallow absorption (Fig.\,\ref{f8}).
No circular polarisation signatures are visible in H$\beta$, H$\gamma$, and 
H$\delta$, but such signatures are clearly observed in the lines of the 
Ca\,{\sc ii} doublet.
The Ca\,{\sc ii} K line shows
a two-component profile which is much wider than predicted by the
synthetic photospheric spectrum. It is very likely that this line is purely of CS origin and forms
in an envelope containing outflowing as well as infalling gas.
No indications of a photospheric magnetic field have been found in HD\,163296.
This object demonstrates a signature of the presence of a magnetic field only in the CS
Ca\,{\sc ii} lines.

\subsubsection{HD\,190073}\label{iiic}

\begin{figure}
\centering
%\sidecaption
\resizebox{.45\textwidth}{!}{\includegraphics{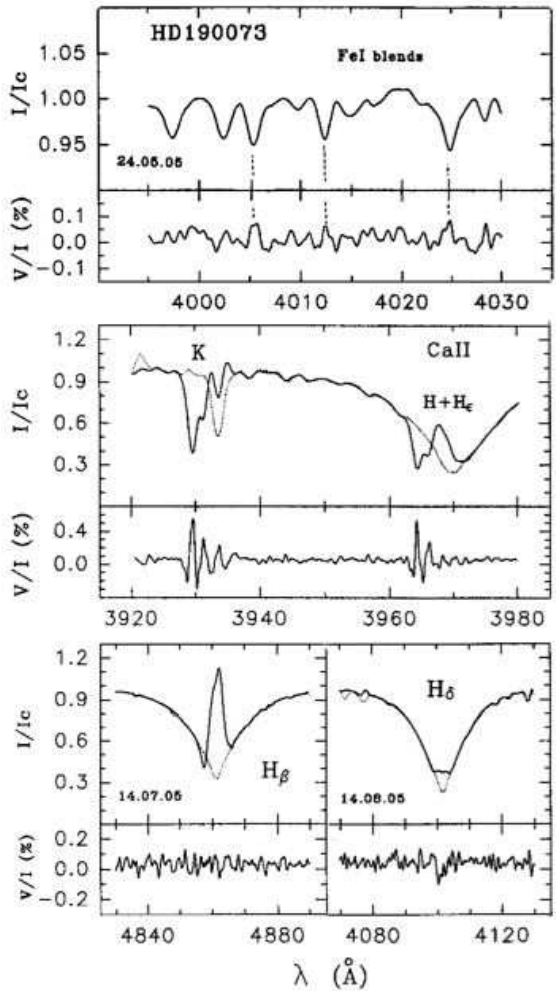}}
%\caption[]{The same caption as in Fig.\,\ref{f1} but for 
\caption[]{Stokes~I and V spectra of HD\,190073 observed on May 24, 2005 (upper panel) 
and August 14, 2005 (bottom). The middle panel presents the mean Ca\,{\sc ii} spectrum.
The synthetic photospheric spectrum is shown in all panels by the dotted line.}
\label{f9}
\end{figure}

The most interesting results have been obtained for HD\,190073 which is
the third object of our programme with P\,Cyg-type emissions in the
Balmer lines. Three spectra of the star have been obtained in three
different nights in 2005 (see Table\,\ref{t2}). This allowed us to
get an idea of the temporal behaviour of the object.
The H$\beta$, H$\gamma$, and H$\delta$ emissions are slightly variable
with the P\,Cyg structure being best developed on August 14, 2005
(see Fig.\,\ref{f9}, bottom). This structure in H$\beta$ looks as of 
P\,Cyg-type\,{\sc ii} with a
rather weak blueshifted absorption. However, in H$\delta$ this absorption 
practically disappears. No circular polarisation signatures are found to be 
associated with these lines.

The most prominent spectral peculiarity of HD\,190073 is the complex
multi-component structure of the Ca\,{\sc ii} H and K absorption profiles.
It consists of about 20 blueshifted components of different width and
depth \citep[][and references therein]{pog}. The structure of the profiles
is variable on timescales from years to decades but remains
practically constant during a year. Our spectropolarimetry carried out at low
spectral resolution allowed to distinguish only the three strongest components.
We do not detect any conspicuous profile variations. The
mean Ca\,{\sc ii} spectrum is presented in the middle panel of
Fig.\,\ref{f9}. Distinctive circular polarisation
signatures are clearly visible, displaying three components in both
Ca\,{\sc ii} H and K lines. Their positions coincide with the
positions of three blueshifted components of the line profiles in the Stokes~I spectrum. 
The polarisation feature corresponding to the unshifted component, possibly of 
photospheric origin, has the lowest amplitude. The highest amplitude is 
observed in the most blueshifted component. This behaviour leads us to the 
conclusion that a 
direct relation must exist between the specific structure of the Ca\,{\sc ii} H and K
line profiles observed in the spectrum of HD\,190073 and the CS magnetic
field.

A search for the photospheric magnetic field in HD\,190073 and the study of 
its temporal behaviour revealed no significant polarisation signatures
in the spectra observed at May 24 and 29, 2005. However, a weak magnetic 
field has been measured on the spectrum obtained in  August 14, 2005 
($\langle$$B_z$$\rangle$\,=\,$-51\pm$14\,G).

\section {Discussion}\label{dis}

One of the important
characteristics of the HAEBEs is the presence of variable CS contribution
to their stellar spectra. This contribution strongly affects the
results of spectropolarimetric studies of magnetic fields in this type of
objects. If the CS signatures are rather weak and observed
only in a limited number of spectral lines, like in the first members of
the Balmer series, then it is possible to choose spectral regions which are 
not contaminated by the 
CS material and to measure the photospheric magnetic field with rather high
precision (objects of group I). For the observations with a stronger CS contribution the field 
detection becomes tough, because of the difficulty
to identify spectral regions with clean photospheric
stellar lines (objects of group II). If CS lines become
dominant in the spectrum of a Herbig star, then there is no way to measure 
the photospheric field, but it is possible to study the magnetic field present
in the CS environment (objects of group III).

As a result of our magnetic field investigation of seven HAEBEs,
%using the method of separation of photospheric and CS magnetic field components,
we make the following {\change assertions}:
\begin{itemize}
\item We conclude that previous discrepancies in estimations of magnetic
fields for a few HAEBEs observed on different dates are likely the result of
a variable CS contribution to photospheric spectra of these objects.
\item We have improved the accuracy of the photospheric magnetic field
determination of the two objects of group I, HD\,139614 and HD\,144432 (observed
in 2005), to a significance level of about 7$\sigma$, instead of the previous 
3$\sigma$ detections.
This leaves little doubt about the presence of a magnetic field.
\item The Stokes~V spectra of all seven programme
targets demonstrate notable circular polarisation signatures, although for 
a few Herbig Ae stars, the effective magnetic field is too weak to be detected
at a high significance level.
\item We deduce that the measured magnetic fields of HD\,31648
and HD\,190073 \citep{h6}  are not of photospheric but mostly (HD\,31648) or exclusively (HD\,190073) of CS origin.
\item From the analysis of polarisation signatures of the CS magnetic field in the
Herbig Ae stars HD\,31648 and HD\,144432
we conclude that in the Balmer lines they are only observed in deep, blueshifted
absorption components 
formed in a dense stellar wind in the region around the star.
%we conclude that they are observed only in deep blueshifted
%absorption components of Balmer lines formed in a dense stellar wind in the region 
%around the stars.
\item We found that the most sensitive indicator of the CS magnetic field
in Herbig Ae stars is the Ca\,{\sc ii} doublet. Circular polarisation features
corresponding to this doublet are observed in all programme targets of
groups II and III.
From the analysis of HD\,31648 and HD\,163296
we suggest that the magnetic field diagnosed in the Ca\,{\sc ii}
lines is that present in the CS matter in the vicinity of the stellar surface where
the base of the stellar wind as well as gaseous flows infalling onto the star
are likely located.
\item {\change The results of our study of HD\,190073 have
confirmed the magnetic nature of the well-known stable
complex structure of absorption Ca\,{\sc ii} profiles in the
spectrum of the object. The observations presented in Fig.\,\ref{f9} are 
in line with the hypothesis of \citet{pog} who suggested that this
structure is a result of latitudinal stratification
of the stellar wind governed by a global magnetic field
with complex topology.}
%An important result has been obtained in the course of our
%spectropolarimetric study of HD\,190073.
%In \citet{pog} a new hypothesis {\change was} put
%forward that the stable complex structure of Ca\,{\sc ii} profiles in the
%spectrum of this object is caused by a latitudinal stratification of
%the stellar wind near the line-of-sight, which remains azimuthally
%homogeneous. In this assumption, each discrete local
%component of the profile is formed in a separate latitudinal layer of the stellar
%wind governed by the magnetic field. Taking into account that a magnetic
%centrifuge is recognized to be one of the principal mechanisms of wind
%acceleration \citep{bl}, we conclude that the velocity of the gaseous outflow
%is directly related to the strength of the field.
%The observations presented in Fig.\,\ref{f9} are clearly in favour of the hypothesis 
%presented by \citet{pog}.
\end{itemize}

We would like to emphasise that the most effective strategy for future
studies of magnetic fields of HAEBEs will be to carry out spectropolarimetric
monitoring of these stars at different states of their CS
contribution to the photospheric spectrum. In the state of minimum CS distortion 
the photospheric magnetic field is relatively easy to detect. When the CS spectrum becomes
the most prominent, one has a good chance of studying the field
in the CS environment.
We also note that spectropolarimetric observations carried out at higher spectral
resolution would allow to analyse polarisation signatures in more detail and to
use a considerably larger sample of spectral lines originating at different levels
in the stellar atmosphere and the CS envelope.
This should provide an opportunity to reconstruct the magnetic field topology from the stellar
surface to the external CS envelope.

\begin{acknowledgements}
{\change We would like to thank the referee, Richard Ignace, for his valuable comments.}
\end{acknowledgements}

%\bibliography{astrobib}

%\bibliographystyle{aa}

\end{document}